\documentclass[twocolumn,aps,showpacs,prc]{revtex4}
\usepackage[dvips]{epsfig}
\begin{document}	

\title{Direct photon production from viscous QGP}
 
\author{A. K. Chaudhuri}
\email[E-mail:]{akc@veccal.ernet.in}
\affiliation{Variable Energy Cyclotron Centre,\\ 1/AF, Bidhan Nagar,
Kolkata 700~064, India}
\author{Bikash Sinha}
\email[E-mail:]{bikash@veccal.ernet.in}
\affiliation{Variable Energy Cyclotron Centre,\\ 1/AF, Bidhan Nagar,
Kolkata 700~064, India}

\begin{abstract}
We simulate direct photon production in evolution of viscous QGP medium. Photons from Compton and annihilation processes are considered. Viscous effect on photon production is very strong and reliable simulation is possible only in a   limited $p_T$ range. For minimally viscous fluid $\eta/s$=0.08), direct photons can be reliably computed only up to $p_T \leq$ 1.3 GeV.   With reduced viscosity  ($\eta/s$=0.04), the limit increases to $p_T \leq $2GeV. 
%We argue that in viscous dynamic, with similar initial conditions, simultaneous fit to experimental data for charged particles and direct photon will not be possible.
 \end{abstract} 

\pacs{ PACS numbers(s):25.75.-q,12.38.Mh} 

\maketitle
\section{Introduction}

Recent experiments at Relativistic Heavy Ion Collider (RHIC) have produced convincing evidences that in $\sqrt{s}_{NN}$=200 GeV Au+Au collisions, a collective medium is produced \cite{BRAHMSwhitepaper,PHOBOSwhitepaper,PHENIXwhitepaper,STARwhitepaper} .
Strong elliptic flow in non-central collisions is key evidence to this
understanding. Hydrodynamical analysis of experimental charged particles data,
also suggests that in Au+Au collisions, 
a collective medium, with viscosity to entropy ratio close to the AdS/CFT lower bound of viscosity,   $\eta/s \geq 1/4\pi$ is produced \cite{QGP3,Chaudhuri:2009uk,Chaudhuri:2008ed}. Whether the collective medium is a deconfined Quark-Gluon Plasma (QGP) or not, is still a question of debate. Hopefully, the issue will be settled in Pb+Pb collisions at the Large Hadron Collider (LHC).

Direct photons probe  the {\em early} medium produced in a collision better than the charged hadrons. 
Hadrons, being strongly interacting, are emitted from the  surface of the thermalised matter and
carry information about the freeze-out surface only. They are unaware of
the condition of the interior of the matter and can provide information about the deep interior only in an indirect way.
In a hydrodynamic model, one fixes the initial conditions of the fluid  such that the "experimental" freeze-out surface is correctly reproduced.  
In contrast 
to hadrons,    photons, being weakly interacting, are emitted from 
whole volume of the matter. Throughout the evolution of the matter, photons are emitted. Conditions of the produced matter, at its deep interior,
are better probed by the photons. Depending on the transverse momentum, direct photons can
probe different aspects of heavy ion collisions. A thermalised
medium of quarks and gluons, or of hadrons, can produce
significant number of thermal photons. They are
low $p_T$ photons ($p_T \leq 3$ GeV/c). Low $p_T$ photons can test whether or not, QGP is produced in Au+Au collisions. Hard  
photons ($p_T >$ 6 GeV/c) are of pQCD origin and test the
pQCD models. Fast partons from 'jet' can interact with 
thermal partons of QGP and produce photons. 
At intermediate $p_T$ range, ($3\leq p_T \leq$ 6 GeV/c), interaction jets with QGP could be an important source of direct photons \cite{Fries:2002kt,Gale:2005zd}.

In ideal hydrodynamic models, 
photon production in Au+Au collisions at RHIC energy has been studied extensively \cite{Alam:2007dv,Chatterjee:2005de}. However, it is now realized that the strongly interacting medium, produced in Au+Au collisions, must be treated as a viscous medium. Gravity dual theories suggest that specific viscosity, i.e. viscosity to entropy ratio of {\em any matter} has a lower bound, the so called KSS bound $\eta/s=1/4\pi$ \cite{Policastro:2001yc,Kovtun:2003wp}. Even though, photons are important probe of QGP matter,
 viscous effects on photon production are not studied much. Only recently, Dusling  \cite{Dusling:2009bc} studied viscous effects on photon production from a QGP medium. However, the model appears to have some inconsistency.  In viscous evolution, photon production is affected due to (i) modified fluid evolution and (ii) non-equilibrium correction to equilibrium distribution function. In \cite{Dusling:2009bc}, while
 the non-equilibrium correction to the distribution function was included,
 modification of the fluid evolution, due to viscosity  was neglected. In the present paper, the inconsistency is removed. %We also make some detail study. 

\section{Hydrodynical equations, equation of state and initial conditions}
 
In a hydrodynamical model, the invariant distribution of direct photons is obtained by convoluting the  photon production rate with space-time evolution of the fluid. We assume that in $\sqrt{s}_{NN}$=200 GeV, Au+Au collisions  at RHIC, a baryon free QGP fluid is formed. Space-time evolution of the fluid is obtained by solving 2nd order Israel-Stewart's theory,

\begin{eqnarray}  
\partial_\mu T^{\mu\nu} & = & 0,  \label{eq1} \\
D\pi^{\mu\nu} & = & -\frac{1}{\tau_\pi} (\pi^{\mu\nu}-2\eta \nabla^{<\mu} u^{\nu>}) \nonumber \\
&-&[u^\mu\pi^{\nu\lambda}+u^\nu\pi^{\nu\lambda}]Du_\lambda. \label{eq2}
\end{eqnarray}

Eq.\ref{eq1} is the conservation equation for the energy-momentum tensor, $T^{\mu\nu}=(\varepsilon+p)u^\mu u^\nu - pg^{\mu\nu}+\pi^{\mu\nu}$, 
$\varepsilon$, $p$ and $u$ being the energy density, pressure and fluid velocity respectively. $\pi^{\mu\nu}$ is the shear stress tensor (we are neglecting bulk viscosity). Eq.\ref{eq2} is the relaxation equation for the shear stress tensor $\pi^{\mu\nu}$.   
In Eq.\ref{eq2}, $D=u^\mu \partial_\mu$ is the convective time derivative, $\nabla^{<\mu} u^{\nu>}= \frac{1}{2}(\nabla^\mu u^\nu + \nabla^\nu u^\mu)-\frac{1}{3}  
(\partial . u) (g^{\mu\nu}-u^\mu u^\nu)$ is a symmetric traceless tensor. $\eta$ is the shear viscosity and $\tau_\pi$ is the relaxation time.  It may be mentioned that in a conformally symmetric fluid relaxation equation can contain additional terms  \cite{Song:2008si}.
Assuming boost-invariance, Eqs.\ref{eq1} and \ref{eq2}  are solved in $(\tau=\sqrt{t^2-z^2},x,y,\eta_s=\frac{1}{2}\ln\frac{t+z}{t-z})$ coordinates, with the code   "`AZHYDRO-KOLKATA"', developed at the Cyclotron Centre, Kolkata.
 Details of the code can be found in \cite{Chaudhuri:2008sj,Chaudhuri:2009uk,Chaudhuri:2008ed}. 
%Within 10\% or less, AZHYDRO-KOLKATA simulation  reproduces  Song and Heinz's  \cite{Song:2008si} result for temporal evolution of momentum anisotropy $\varepsilon_p$.

 Eqs.\ref{eq1},\ref{eq2} are closed with an equation of state $p=p(\varepsilon)$.
Lattice simulations      \cite{Cheng:2007jq,Cheng:2009zi,Aoki:2006br,Aoki:2009sc}
indicate that the confinement-deconfinement transition is a cross over, rather than a 1st or 2nd order phase transition. There is no critical temperature for a cross-over transition. However, one can define a pseudo critical temperature, the inflection point on the Polyakov loop. In Wuppertal-Budapest simulation \cite{Aoki:2006br,Aoki:2009sc}, pseudo critical temperature 
$T_c\approx$170 MeV.  
In the present simulation, for the QGP phase, we use an EOS based on Wuppertal-Budapest simulation.

Solution of partial differential equations (Eqs.\ref{eq1},\ref{eq2}) requires initial conditions, e.g.  transverse profile of the energy density ($\varepsilon(x,y)$), fluid velocity ($v_x(x,y),v_y(x,y)$) and shear stress tensor ($\pi^{\mu\nu}(x,y)$) at the initial time $\tau_i$. One also need to specify the viscosity ($\eta$) and the relaxation time ($\tau_\pi$). A freeze-out
temperature is also needed. In the following, we will consider viscous effects on photons from the QGP phase only. The hydrodynamical equations are then solved till the freeze-out temperature $T_F=T_c$=170 MeV. At the initial time $\tau_i$, initial energy density is assumed to be distributed as \cite{QGP3}

\begin{equation} \label{eq6}
\varepsilon({\bf b},x,y)=\varepsilon_i[(1-f_{hard} N_{part}({\bf b},x,y) +f_{hard} N_{coll}({\bf b},x,y)],
\end{equation}

\noindent
where b is the impact parameter of the collision. $N_{part}$ and $N_{coll}$ are the transverse profile of the average participant and collision number respectively, calculated in a Glauber model. $f_{hard}$=0.13 is the hard scattering fraction \cite{Hirano:2009ah}. 
$\varepsilon_i$ is the central energy density of the fluid in impact parameter $b=0$ collision. As it will be discussed later, we have simulated Au+Au collisions for a range of initial (central) energy density and initial time.
We also assume that initial fluid velocity is zero, $v_x(x,y)=v_y(x,y)=0$. The shear stress tensor was initialized with boost-invariant value, $\pi^{xx}=\pi^{yy}=2\eta/3\tau_i$, $\pi^{xy}$=0. For the relaxation time, we use the Boltzmann estimate $\tau_\pi=3\eta/4p$. We also assume that shear viscosity to entropy ratio is a constant throughout the evolution. In the following, we simulate Au+Au collisions for $\eta/s$=0-0.12. 

\section{photon rates}

As mentioned earlier,  in viscous evolution photon rates are modified. The photon rate equations  involve distribution functions of quarks and gluons. For example, in $1+2\rightarrow 3+4$ processes (e.g. Compton and annihilation processes), the general form of photon production rate is,

\begin{widetext}
\begin{equation}
E_4\frac{dR}{d^3p_4}=\mathcal{N}\int \frac{d^3p_1}{(2\pi)^3 2E_1} \frac{d^3p_2}{(2\pi)^3 2E_2} 
 f_1(E1)f_2(E_2)
(2\pi) \delta^4(p_1+p_2-p_3-p_4)|M|^2 [1\pm f_3(E_3)]\frac{d^3p_3}{(2\pi)^3 2E_3}  \label{eq4}
\end{equation}
\end{widetext}

\noindent where $M$ is the matrix element for the reaction. 
$f$'s in Eq.\ref{eq4} are distribution function of quarks and gluons.
Unlike an ideal fluid,   
in viscous evolution, each distribution functions $f$ is modified due to non-equilibrium correction,

\begin{equation}
f_{neq} \rightarrow f_{eq}(1+\delta f_{neq}), \delta f_{neq} << 1.
\end{equation}

The non-equilibrium correction $\delta f_{neq}$ depend on dissipative forces as well as on particle momenta. For shear viscosity, it can be obtained as,

\begin{equation} \label{eq6}
\delta f_{neq}=C p_\mu p_\nu \pi^{\mu\nu}=\frac{1}{2(\varepsilon+p)T^2}p_\mu p_\nu \pi^{\mu\nu}
\end{equation}

It is obvious that non-equilibrium correction to photon rates is non-trivial.
For Compton and annihilation processes, in the leading log approximation, the rate equation can be simplified \cite{Dusling:2009bc},

\begin{equation}
E\frac{dR}{d^3p}\approx \frac{2}{(2\pi)^6} f_1(E) \int d^3p_2 f_2(E_2)
[1\pm f_3(E_2) \frac{s\sigma(s)}{E_2}
\end{equation}

Non-equilibrium correction to distribution functions inside the phase space integral leads to corrections in the logarithmic term and can be neglected. For equilibrium distribution function, as demonstrated by Kapusta and Lichard \cite{Kapusta:1991qp}, the phase integration can be explicitly evaluated and one obtain for both Compton and annihilation processes \cite{Dusling:2009bc},
 
\begin{equation} \label{eq8}
E\frac{dR}{d^3p}\approx \frac{5}{9}\frac{\alpha_e \alpha_s}{2\pi^2} f_{neq}(E)
T^2 \ln \left [ \frac{3.7388E}{g^2T}\right ]
\end{equation} 

The invariant photon distribution then has two parts, equilibrium part
($E\frac{dN_{eq}}{d^3p}$)  and a non-equilibrium part ($E\frac{dN_{neq}}{d^3p}$).  Since non-equilibrium correction to distribution function is assumed to be small, it is essential that,

\begin{equation}
E\frac{dN_{neq}}{d^3p} << E\frac{dN_{eq}}{d^3p}.
\end{equation}

%Later, we will show that the condition restrict the applicability of viscous hydrodynamics only in a limited $p_T$ range for photon production from QGP fluid.

\begin{figure}[t]
%\vspace{0.3cm} 
\center
\resizebox{0.4\textwidth}{!}{%
  \includegraphics{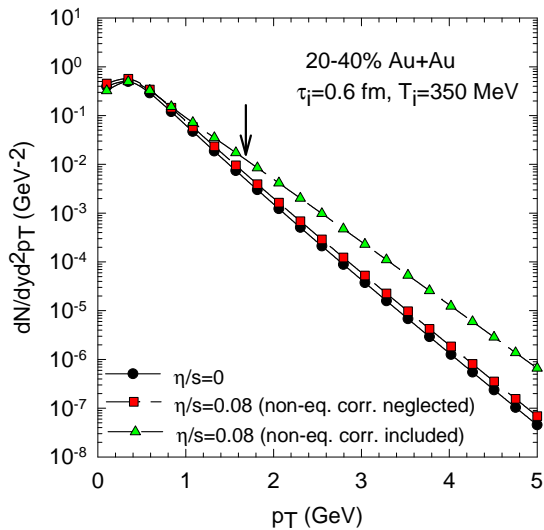}
}
%\vspace{-4cm}
\caption{(color online)
Transverse momentum spectra for photons, from evolution of ideal and minimally viscous ($\eta/s$=0.08) QGP. For viscous fluid, spectra obtained with and without the  non-equilibrium correction to distribution function, are shown separately.
 }\label{F1}
\end{figure}    

\section{Effect of viscosity on photon spectra and elliptic flow}

Gravity dual theories \cite{Policastro:2001yc} indicate that viscosity to entropy ratio of strongly interacting matter is bounded from the lower side, $\eta/s \geq 1/4\pi$. We first consider effect of minimal viscosity on photon production. 
We assume that minimally viscous ($\eta/s$=0.08) QGP fluid is thermalised in the time scale $\tau_i$=0.6 fm to initial central temperature $T_0$=350 MeV.
A large number of
charged particle's data e.g. identified particles spectra, elliptic flow etc.  data are explained in hydrodynamical model with similar initial time and temperature scale \cite{QGP3}. 
In Fig.\ref{F1}, simulated photon spectra in 20-40\% Au+Au collisions, from evolution of ideal and minimally viscous QGP fluid, are shown.
The spectra with or without the  non-equilibrium correction are shown separately.
If non-equilibrium correction to distribution function is neglected, 
$p_T$ spectra of photons is increased by a factor of $\sim$ 1.2-1.5. The increase is largely $p_T$ independent. In contrast, when non-equilibrium correction is included, photon production is increased more at large $p_T$ than at low $p_T$,
  It is also expected, non-equilibrium correction increases with $p_T$ (see Eq.\ref{eq6}). The arrow in Fig.\ref{F1} indicate the approximate $p_T$ when non-equilibrium contribution ($\delta N_{neq}$) to
photon spectra equals the equilibrium contribution $N_{eq}$).  For minimally viscous fluid, the equality occur at $p_T\approx$1.7 GeV. As noted earlier, viscous 
hydrodynamics is applicable only when $\delta N_{neq}/N_{eq} << 1$. Evidently,
for $\eta/s$=0.08,  in a hydrodynamic model, photon production from Compton and annihilation processes can not be reliably computed
beyond $p_T$=1.7 GeV. Indeed, if we assume that viscous hydrodynamics remain reliable only until, $\frac{\delta N_{eq}}{N_{eq}}=0.5$, the photons from Compton and annihilation processes can be computed only up to $p_T\approx$1.3 GeV. 

\begin{figure}[t]
%\vspace{0.3cm} 
\center
\resizebox{0.40\textwidth}{!}{%
  \includegraphics{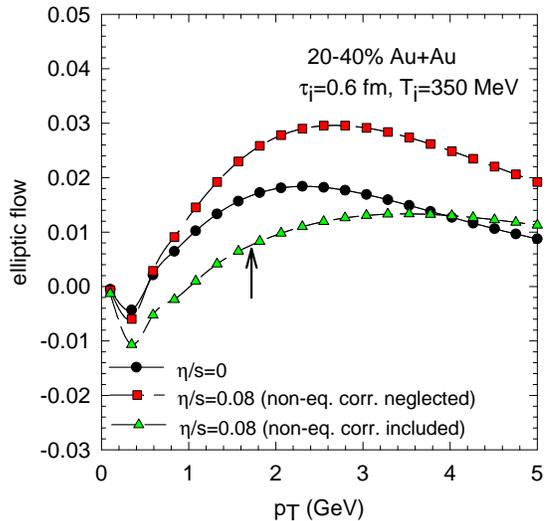}
}
%\vspace{-4cm}
\caption{(color online) same as in Fig.\ref{F1} but for elliptic flow.}\label{F2}
\end{figure}     

In Fig.\ref{F2}, we have shown the simulation results for   elliptic flow for photons. At very low $p_T$ elliptic flow is negative.  At low $p_T \leq 0.5 GeV$   present model is not reliable. The logarithmic factor in the photon rate equation is negative when $E<0.27g^2T$. For fluid (central) temperature $T_0$=350 MeV, reliable computation is possible only beyond $p_T =$0.5 GeV. In ideal fluid evolution, at $p_T>$0.5 GeV, elliptic flow increases with $p_T$, till a maxima 
is reached at $p_T\approx$2 GeV. At even higher $p_T$, it decreases again. Note that even the highest $v_2(p_T)$ is not large, less that $\sim$ 2\%. Compare this value to $v_2(p_T)\sim$ 20\% for charged particles.   In viscous evolution, elliptic flow is reduced further. Incidentally, very small photon elliptic flow is consistent with experiments. In experiments also, photons do not flow \cite{Adler:2005ig,Sahlmueller:2006uk}. 
In Fig.\ref{F2}, one observes that if non-equilibrium correction is neglected, elliptic flow in viscous evolution is more than flow in ideal fluid evolution. However, with non-equilibrium correction included, flow is reduced in viscous evolution. It shows that it is important to have a consistent model, otherwise, one can conclude wrongly. The arrow in Fig.\ref{F2} indicate that approximate
$p_T$ when $\delta N_{neq}\approx N_{eq}$.  
 
\begin{figure}[t]
%\vspace{0.3cm} 
\center
\resizebox{0.40\textwidth}{!}{%
  \includegraphics{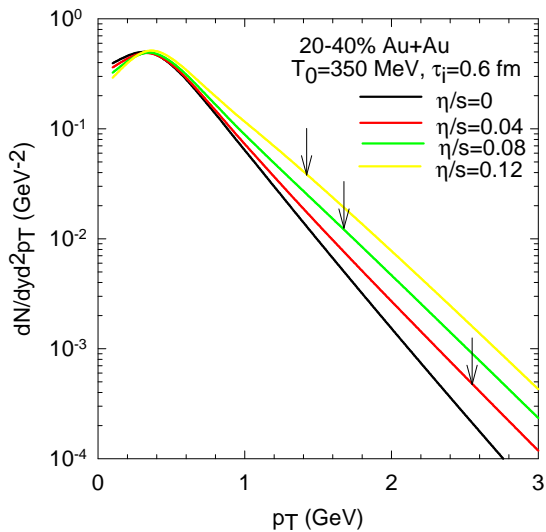}
}
%\vspace{-4cm}
\caption{(color online) Photon spectra from evolution of QGP fluid, with viscosity to entropy ratio, $\eta/s$=0, 0.04, 0.08 and 0.12 are shown. The initial time is $\tau_i$=0.6 fm and initial central temperature is $T_i$=350 MeV.
}\label{F3}
\end{figure}

Photon spectra from Compton and annihilation processes,  for four values of viscosity to entropy ratio, $\eta/s$= 0 (ideal fluid), 0.04, 0.08 (AdS/CFT lower bound) and 0.12 are shown in Fig.\ref{F3}. Initial time and temperatures are $\tau_i$=0.6 fm and $T_i$=350 MeV. As expected, high $p_T$ yield is increased in more viscous fluid evolution. 
%However, effect of viscosity is much larger for photons than for charged particles. 
The arrows in Fig.\ref{F3} indicate the approximate $p_T$ when  $\delta N_{neq}=N_{eq}$.
Photon production can not be computed reliably beyond $p_T$=1.3, 1.7 and 2.4 GeV, for QGP viscosity, $\eta/s$=0.12, 0.08 and 0.04 respectively. Very limited $p_T$ range over which viscous hydrodynamics remain applicable for photon production raises certain issues, which will be discussed later. We just mention that viscous effect on photon production is much stronger than in charged particles.
Indeed, for charged particles, viscous hydrodynamics remain applicable over a much wider $p_T$ range \cite{Chaudhuri:2009uk,Chaudhuri:2008ed}. The reason is understood. Photons are emitted through out the evolution. At early time, shear stress tensors have finite values and non-equilibrium correction to photon production is large. In contrast, charged    particles are emitted from the freeze-out surface. Shear stress tensors evolves very fast. Compared to early times, stress tensors at freeze-out are much smaller in magnitude. Accordingly, non-equilibrium correction is small. We have not shown simulation results for elliptic flow as a function of viscosity.  As shown earlier, $v_2(p_T)$ for photons is very small in ideal fluid evolution. It further reduces with  $\eta/s$.

Since photons are emitted throughout the evolution, including the earliest phase of QGP, one hope to extract QGP formation time from photon measurements. 
Indeed, in \cite{Chatterjee:2009qz,Chatterjee:2009ys}
it was suggested that photon elliptic flow can be used to constrain QGP formation time. However, from the experiments \cite{Adler:2005ig,Sahlmueller:2006uk}, it appears that photons do not seem to experience flow. One can not possibly
extract QGP formation time from elliptic flow measurements.  
In \cite{Dusling:2009bc}, Dusling suggested that inverse slope of the photon spectra, if measured accurately, can be used to put stringent bound on QGP viscosity and initial (formation) time.  

 \begin{figure}[t]
%\vspace{0.3cm} 
\center
\resizebox{0.4\textwidth}{!}{%
  \includegraphics{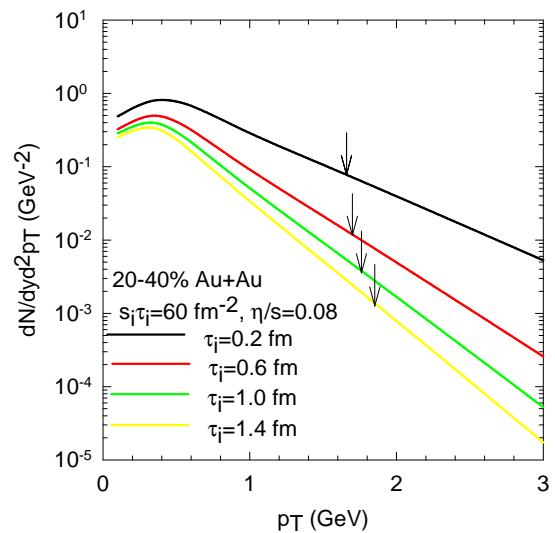}
}
%\vspace{-4cm}
\caption{(color online) Initial time dependence of photon spectra from evolution of QGP fluid with viscosity $\eta/s$=0.08. The initial entropy density times the initial time is fixed
at $s_i\tau_i$=60 $fm^{-2}$.}\label{F4}
\end{figure}   

In Fig.\ref{F4}, simulated photon spectra from evolution of minimally viscous ($\eta/s$=0.08) QGP fluid, for four values of initial time $\tau_i$=0.2, 0.6, 1.0 and 1.4 fm are shown.  The initial temperature is obtained from the condition that initial entropy density times the initial time is a constant, $s_i\tau_i=60 fm^{-3}$.
As the initial time is reduced, photon spectra are hardened. Photon yield is also increased. It is easily understood,  for fixed $s_i\tau_i$, initial time and temperature are inversely related. As the initial time is reduced, high $p_T$ photon production is increased due to increased fluid temperature. 
The arrows in Fig.\ref{F4}, indicate the $p_T$ when non-equilibrium correction to spectra is equal to the equilibrium contribution. 

\begin{figure}[t]
%\vspace{0.3cm} 
\center
\resizebox{0.4\textwidth}{!}{%
  \includegraphics{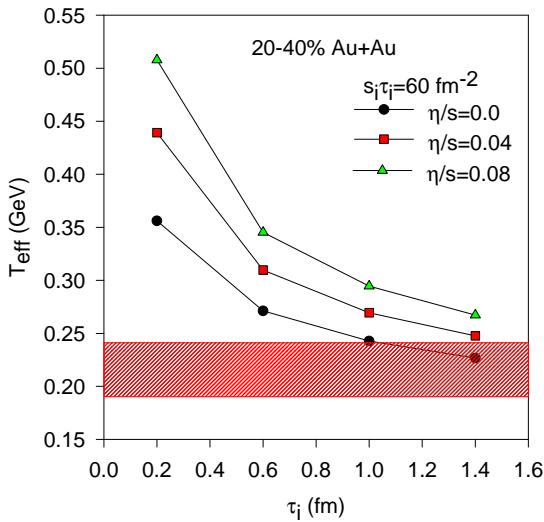}
}
%\vspace{-4cm}
\caption{(color online) Effective temperature from photon spectra for initial time $\tau_i$=0.2, 0.6, 1.0 and 1.4 fm. The initial entropy density times the initial time is fixed
at $s_i\tau_i$=60 $fm^{-2}$. Results are shown for viscosity to entropy ratio 
$\eta/s$=0, 0.04 and 0.8 respectively. The two dashed lines indicate the experimental uncertainty in PHENIX measurements.}\label{F5}
\end{figure}

To obtain the inverse slope parameter $T_{eff}$,  we have fitted 
the spectra in the $p_T$ range $1.5 \leq p_T \leq 2.5$GeV with an exponential   ($dN/d^2p_T \propto e^{-p_T/T_{eff}}$ ). Ideal hydrodynamic simulations for photon spectra suggest that in this $p_T$ range, QGP photons dominate the spectra \cite{Alam:2007dv,Chatterjee:2005de}. 
%However, slope of the photon spectra from QGP is fairly constant for $p_T>$0.5 GeV, and effective temperature will remain essentially unchanged even in a smaller interval. 
In Fig.\ref{F5}, for fluid viscosity, $\eta/s$=0, 0.04 and 0.08, inverse slope parameter is shown as a function of initial time.   Inverse slope parameter $T_{eff}$ decreases with
increasing initial time, as well as with decreasing viscosity. 
%Extracting QGP formation time from inverse slope of the photon spectra will not be easy. Less viscous QGP with small formation time could be confused with more viscous QGP with large formation time.  
In Fig.\ref{F5}, the shaded region indicates the experimental slope measured in the PHENIX experiment. For ideal fluid, simulated $T_{eff}$ agree with experiment for $\tau_i\geq$ 1 fm. For viscous fluid, in the time scale,
$0.2-1.0$ fm, inverse slope of the simulated spectra are higher than that observed in experiment. However, we have neglected photons from the hadronic phase.   Hadronic photons will dominantly  contribute at low $p_T$,
reducing the inverse slope parameter. In other word, in more realistic simulation,
$T_{eff}$ could be smaller than obtained presently and even for  viscous fluid could be in agreement with experiment. 

To constrain initial time and viscosity from inverse slope of photon spectra will not be an easy task. As seen in Fig.\ref{F5}, inverse slope parameter   from evolution of low viscosity fluid initialized at
small $\tau_i$ could be confused with $T_{eff}$ from high viscosity fluid initialised at large $\tau_i$. As shown in Fig.\ref{F5}, for $\eta/s=0.04 \pm 0.04$, depending on the initial time, change in inverse
slope parameters is $\sim$20-80 MeV. 
Depending on the realistic situation, measurement of inverse slope parameter with accuracy $\delta T_{eff}\approx$ 20-80 MeV could possibly give
a range of values for initial time and viscosity.

Before we summarise, few comments are in order. We have shown that effect of viscosity is quite strong on photon production. Indeed, viscous hydrodynamics in applicable only in a limited $p_T$ range. For minimally viscous  fluid, depending on the initial time, applicability range is less than $p_T$=1.5 GeV. 
Applicability range is even less in more viscous fluid.
   Present simulations neglect Bremsstrahlung photons in the QGP phase. However, their inclusion will not improve the situation. One wonders about the photons  in the $p_T$ range $1.5 \leq p_T \leq 3$ GeV. As noted earlier, ideal hydrodynamics simulations indicate that in this $p_T$ range, thermal photons dominate. However, present simulations 
indicate that if QGP fluid is minimally viscous,  the photons in the $p_T$ range $1.5 \leq p_T \leq 3$ GeV can not be considered thermal and can not be described hydrodynamically. Do we understand them as pQCD photons? Or are they from a fluid with viscosity much less than the AdS/CFT lower bound? Indeed, in certain gravity dual theories, KSS bound can be violated by $\sim$36\% \cite{Brigante:2008gz}. Limited $p_T$ range over which photons can be reliably computed in viscous dynamics also indicate that within a same model, one possibly can not explain, charged particle production and photon production.

\section{Summary}

To summarize, we have studied effect of shear viscosity on   Compton and annihilation photons. 
In viscous dynamics, photon production is modified due to (i) changed space-time evolution of the fluid and (ii) non-equilibrium correction to the equilibrium distribution function. 
The non-equilibrium correction grows with viscosity as well with transverse momentum. Viscous effects on photon production are strong. Even for AdS/CFT lower bound of viscosity ($\eta/s$=0.08), strong viscous correction render the hydrodynamics inapplicable beyond $p_T\approx$1.5 GeV. 
 If QGP viscosity is larger than the Ads/CFT limit, then limitation will be even more. Photon production as a function of initial time, also suggest that
 if the   inverse slope parameter of the photon spectra,
 is measured within an accuracy of $\pm$20 MeV, one can possibly limit the
 initial time and viscosity.

   %Charged particles on the other hand can be reliably estimated over much wider $p_T$ range. Limited $p_T$ range over which viscous hydrodynamics remain  applicable raises some interesting questions, e.g. do photons in the $p_T$ range $1.5\leq p_T \leq 3$GeV are thermal, or of pQCD origin? Do  It appears that simultaneous explanation of photon and charged particle production can not be obtained in a hydrodynamic model, unless QGP viscosity is much less than the AdS/CFT lower bound.

\end{document}